\documentclass[runningheads]{llncs}
\usepackage[T1]{fontenc}
\usepackage{comment}
\usepackage{amsmath}
\usepackage{cite}
\usepackage{graphicx}
\begin{document}
\title{Fine-Tuning TransMorph with Gradient Correlation for Anatomical Alignment}
\titlerunning{Gradient Correlation for Anatomical Alignment}
%
\author{Lukas Förner\inst{1,2} \and
Kartikay Tehlan\inst{1,2} \and
Thomas Wendler\inst{1,2,3}}
\authorrunning{L. Förner et al.}
%
\institute{Clinical Computational Medical Imaging Research, Department of Diagnostic and Interventional Radiology and Neuroradiology, University Hospital Augsburg, Augsburg, Germany\\
\email{\{lukas.foerner,kartikay.tehlan,thomas.wendler\}@med.uni-augsburg.de} \and
Computer-Aided Medical Procedures and Augmented Reality, Technical University of Munich, Garching near Munich, Germany \and
Institute of Digital Medicine, University Hospital Augsburg, Neus\"a\ss, Germany}
\maketitle              
\begin{abstract}
Unsupervised deep learning is a promising method in brain MRI registration to reduce the reliance on anatomical labels, while still achieving anatomically accurate transformations. For the Learn2Reg2024 LUMIR challenge, we propose fine-tuning of the pre-trained TransMorph model to improve the convergence stability as well as the deformation smoothness. The former is achieved through the FAdam optimizer, and consistency in structural changes is incorporated through the addition of gradient correlation in the similarity measure, improving anatomical alignment. The results show slight improvements in the Dice and HdDist95 scores, and a notable reduction in the NDV compared to the baseline TransMorph model. These are also confirmed by inspecting the boundaries of the tissue. Our proposed method highlights the effectiveness of including Gradient Correlation to achieve smoother and structurally consistent deformations for interpatient brain MRI registration.

\end{abstract}
\section{Motivation}
Deep-learning approaches to brain magnetic resonance imaging (MRI) registration present challenges to achieving realistic anatomical transformations without relying on predefined anatomical labels. Labeled images help produce high segmentation scores, yet often struggle with smoothness and generalizability across unseen data, leading to deformations that lack biological plausibility \cite{dorent_learn2reg_2024}. Given the availability of expansive neuroimaging datasets, to address these needs in Learn2Reg'24 intersubject T1-weighted brain MRI registration challenge, we propose an unsupervised algorithm tailored to retain structural alignment and minimize deformation inconsistencies. We have focused on fine-tuning the best-performing baseline models to avoid unnecessary re-training, and minimizing the environmental impact. Our approach incorporates gradient correlation for consistent changes in structure and preservation of local structural continuity across subjects. Ensuring alignment of similar structural gradients supports anatomical coherence and enables smoother deformations.

\section{Methods}
To address the Learn2Reg LUMIR challenge, we built on the TransMorph \cite{chen_transmorph_2022} model, for which weights were kindly provided as a baseline model by the challenge organizers. Our main idea consisted of fine-tuning the pre-trained weights of the TransMorph model to help it achieve better results in this unsupervised learning task.

\subsection{FAdam}
One of our modifications was to employ the Fisher Adam (FAdam) \cite{hwang_fadam_2024} optimizer to improve optimization during training. FAdam is a variant of the Adam optimizer that incorporates principles from natural gradient descent and Riemannian geometry to improve convergence and stability. The key modifications proposed by the authors in FAdam include adjustments to momentum, bias correction, and adaptive gradient scaling, alongside a refined weight decay that respects the geometry of the parameter space. FAdam’s superior performance has been demonstrated across multiple domains (language, speech, and image tasks).

\subsection{Normalized Gradient Cross Correlation}
Our second modification is the incorporation of the gradient correlation (GC) \cite{penney_comparison_1998,hiasa_cross-modality_2018} similarity measure. For two three-dimensional images A and B, GC is defined as
\begin{equation}
\text{GC}(A, B) = \frac{1}{3} \left( \text{NCC}(\nabla_x A, \nabla_x B) + \text{NCC}(\nabla_y A, \nabla_y B)+ \text{NCC}(\nabla_z A, \nabla_z B) \right)
\end{equation}
where $\nabla_x$, $\nabla_y$, $\nabla_z$ denotes the gradient in directions x, y, z and $\text{NCC}$ is the normalized cross-correlation given by
\begin{equation}  
\text{NCC}(A, B) = \frac{\sum_{(i,j)} (A - \bar{A})(B - \bar{B})}{\sqrt{\sum_{(i,j)} (A - \bar{A})^2} \sqrt{\sum_{(i,j)} (B - \bar{B})^2}}
\end{equation}
We define the gradient correlation loss $\mathcal{L}_{GC}$ as
\begin{equation}  
\mathcal{L}_{GC} = 1 - \text{GC}(I_t, I_s\circ \phi)
\end{equation}
with $I_t$ being the fixed image, $I_s$ the moving, and $\phi$ the deformation field, and extend the image similarity loss to
\begin{equation}  
\mathcal{L}_{sim} = \mathcal{L}_{IC}(I_t, I_s\circ \phi) + \gamma \mathcal{L}_{GC}(I_t, I_s\circ \phi)
\end{equation}
with $\gamma = 0.5$ to have about equal values between $\mathcal{L}_{IC}$, which is the local normalized cross-correlation (LNCC) loss on the intensity values, and $\mathcal{L}_{GC}$. Our final objective function is given by
\begin{equation}  
\mathcal{L} = \mathcal{L}_{\text{sim}}(I_t, I_s \circ \phi) + \lambda \mathcal{L}_{\text{reg}}(\phi)
\end{equation}
where $\lambda = 2$ to account for the similarity loss being now about double compared to the baseline, and $\mathcal{L}_{\text{reg}}$ being the diffusion regularizer used in TransMorph.
\section{Experiments and Results}
We evaluated our approach using data from the 2024 Learn2Reg LUMIR challenge. The dataset consists of over 4,000 preprocessed T1-weighted brain MRI scans from multiple public sources, including OpenBHB \cite{dufumier_openbhb_2022} and AFIDs \cite{taha_magnetic_2023} (which uses data from OASIS \cite{marcus_open_2007}). All images were given in the NifTi format with a resolution of $160\times 224\times 192$ and a voxel spacing of $1\times 1\times 1 \text{ mm}^3$.
\subsection{Quantitative evaluation}
The performance is evaluated by calculating the segmentation accuracy consisting of Dice and the $95\%$ Hausdorff distance (HdDist95 $\uparrow$). Landmark accuracy is measured using the target registration error (TRE $\downarrow$) of manually annotated landmarks. The final measure is the deformation smoothness, which is quantified by non-diffeomorphic volume (NDV $\downarrow$) \cite{liu_finite_2024}.

We evaluated the performance of three models - ``Baseline'' (which is the TransMorph model provided by the Learn2Reg challenge organizers), ``FAdam'' (which is the TransMorph provided fine-tuned with FAdam) and ``FAdam+GC'' (which is the TransMorph provided fine-tuned with FAdam and GC). For our fine-tuned models, we set the initial learning rate to half of the learning rate used for training the baseline model and used the same scheduler. Both models were trained for an additional 200 epochs and using the version with the lowest evaluation loss at 123 and 170 epochs, respectively.

\begin{table}[h!]
    \centering
    \begin{tabular}{|c|c|c|c|c|}
        \hline
        \textbf{Model}  & \textbf{Dice $\uparrow$ }& \textbf{ TRE (mm) $\downarrow$ } & \textbf{ HdDist95 $\downarrow$ } & \textbf{ NDV (\%) $\downarrow$ } \\ \hline
        Baseline  & $ 0.7594 \pm 0.0319 $ & \textbf{2.4225} & 3.5074 & 0.3509 \\ \hline
        FAdam  & $ 0.7597 \pm 0.0307 $ & 2.4590 & 3.5233 & 0.3549 \\ \hline
        FAdam+GC  & \textbf{ 0.7614 $ \pm $ 0.0316 } & 2.4599 & \textbf{3.4899} & \textbf{0.2690} \\ \hline
    \end{tabular}
    \caption{Evaluation metrics for different models}
    \label{tab:metrics}
\end{table}

As can be seen in table \ref{tab:metrics}, all models perform quite similar in terms of the Dice coefficient with ``FAdam+GC'' slightly outperforming the two other models. The TRE ranges from 2.42 to 2.46 mm indicating comparable registration performance among the evaluated models. HdDist95 shows again minor improvements of ``FAdam+GC'', reducing the error from about 3.5 to 3.4. NDV shows the most notable difference of all evaluated metrics, with ``FAdam+GC'' achieving substantially lower percentage (0.27\%) compared to baseline (0.35\%), suggesting smoother deformations. 

\subsection{Qualitative evaluation}
\begin{figure}
\includegraphics[width=\textwidth]{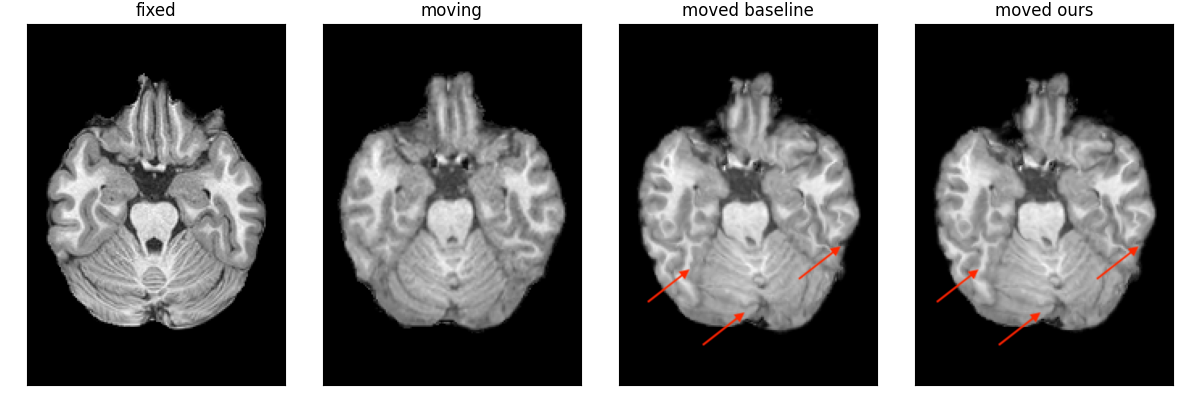}
\caption{Moving, fixed, moved (``Baseline'') and moved (``FAdam+GC'') images. Arrows indicate some example areas where the smoothness of the deformation was improved.} \label{fig:im_comp}
\end{figure}

\begin{figure}
\includegraphics[width=\textwidth]{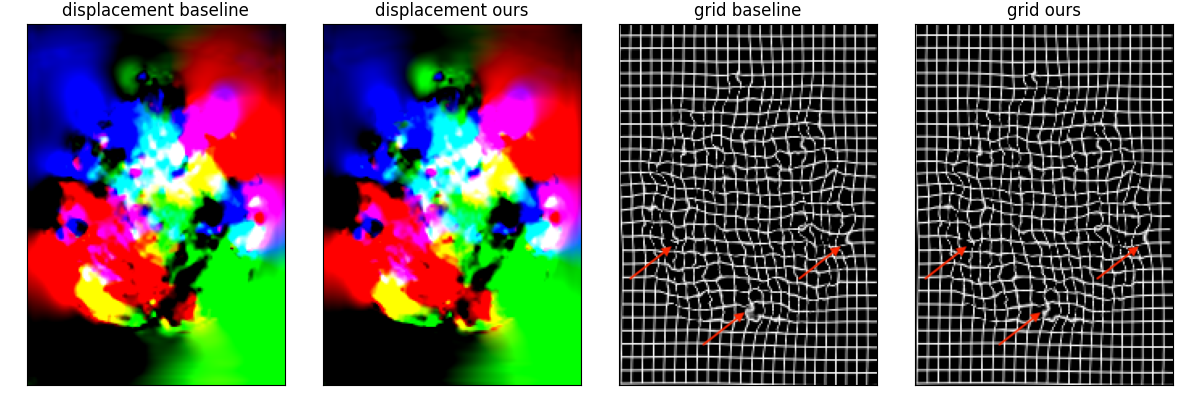}
\caption{Displacement fields and grids of ``Baseline'' and ``FAdam+GC'' - as RGB image coding XYZ displacements (left) and as grid (right). The arrows indicate some example areas where the smoothness of the deformation was improved (cfg. figure \ref{fig:dis_grid}).} \label{fig:dis_grid}
\end{figure}

For a qualitative comparison, we investigated the difference between ``Baseline'' compared to ``FAdam+GC''. For this comparison, image numbers 3457 and 3456 of the LUMIR dataset are registered with both models, and slice number 70 is analyzed. The compared slice can be seen in figure \ref{fig:im_comp} and the corresponding deformation fields are shown in figure \ref{fig:dis_grid}. 
 
\begin{figure}
\includegraphics[width=\textwidth]{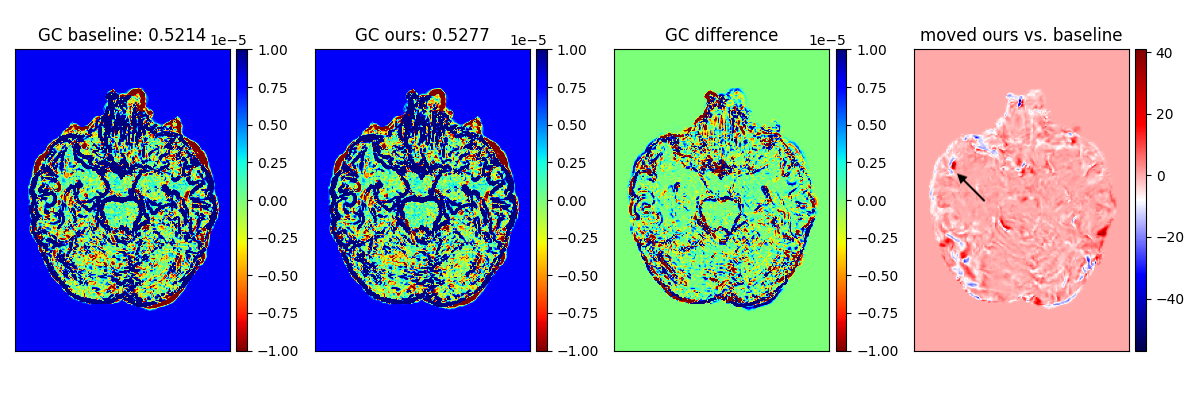}
\caption{GC values of ``Baseline'' and ``FAdam+GC'' as well as the difference in GC and intensity values'. The arrow indicates an example of an area where an artifact created by the baseline was reduced.} \label{fig:gc_comp}
\end{figure}

As can be seen in Figure \ref{fig:gc_comp} our method improves the gradient correlation (0.521 in baseline vs 0.528 in ours). This leads to better results in the evaluation metrics. The main differences are seen at the tissue boundaries where the gradient is high. This indicates a better alignment of the boundaries which is reflected by the metric HdDist95. Another major difference in the intensities is highlighted in the regions where the baseline created strong anatomical structures originally absent in the moving image - such as in the right superior temporal gyrus, whereas this artifact is weaker in our method.

\section{Conclusion}
Our proposed method builds upon the baseline TransMorph model, leveraging fine-tuned pre-trained weights to enhance unsupervised brain MRI registration. The model incorporates the Fisher Adam (FAdam) optimizer, a variant of the Adam optimizer with concepts from natural gradient descent and Riemannian geometry, to improve convergence stability. The integration of gradient correlation (GC) as a similarity measure aims to maintain anatomical alignment by promoting consistent structural changes and preserving local continuity across subjects.

Quantitative evaluation on the 2024 Learn2Reg LUMIR challenge data shows that the proposed \emph{FAdam+GC} model achieves slight improvements in the Dice coefficient and 95\% Hausdorff distance (HdDist95) relative to the baseline. Notably, it demonstrates a reduced percentage of non-diffeomorphic volume (NDV), indicating smoother, anatomically plausible deformations. Qualitative analysis also shows enhanced alignment at tissue interfaces with high gradient values, directly impacting HdDist95 metrics. Our proposed method shows the effectiveness of integrating GC in achieving smoother, structurally consistent deformations for interpatient brain MRI registration. Further, reviews from medical experts of the achieved transformed moved images would be highly beneficial in evaluating the anatomical accuracy of our registration method.

\subsubsection*{Acknowledgement.}
This research was partially funded by the Bavarian Ministry of Economic Affairs, Regional Development and Energy (StMWi) under grant number DIK-2310-0004// DIK0556/02.

%
%
%
 \bibliographystyle{splncs04}
 \bibliography{Learn2Reg2024}

\end{document}